# The Effects of Rolling Deformation and Annealing Treatment on Damping Capacity of 1200 Aluminium Alloy

*M. N. Mazlee[1,a], J. B. Shamsul[1,b], Y. Yasmin[2,c],

S. R. Shamsudin[2,d], M. S. Risby[3,e], M. Afendi[4,f]

[1]Sustainable Engineering Cluster, School of Materials Engineering,

Universiti Malaysia Perlis, 01000 Kangar, Perlis, MALAYSIA

[2]School of Materials Engineering, Universiti Malaysia Perlis,

02600 Arau, Perlis, MALAYSIA

[3]Faculty of Engineering, Universiti Pertahanan Nasional Malaysia (UPNM)

Kem Sg. Besi, 57000 Kuala Lumpur, MALAYSIA

[4]School of Mechatronics Engineering, Universiti Malaysia Perlis,

Pauh, 02600 Arau, Perlis, MALAYSIA

[a]mazlee@unimap.edu.my, [b]sbaharin@unimap.edu.my, [c]yasminyuriz@yahoo.com,
[d]rizam@unimap.edu.my, [e]risby@upnm.edu.my, [f]affendirojan@unimap.edu.my

**Abstract.** Annealing treatment is an important step of rolling deformation that contributes to microstructural evolution and leads to the significant changes in damping capacity. Damping capacities were analyzed in the parallel to rolling direction at 1 and 10 Hz respectively. It was found that severe plastic deformation at 40 percent reduction has lower damping capacity compared to that of 30 percent and 20 percent reductions respectively. The microstructural results show that the grains of as rolled alloys were changed to almost equiaxed structures after a rolling reduction at 40 percent reduction.

**Introduction**

Progress in technology and industry is based on developments in materials and the related heat treatment processes involved. The damping capacity of a material is determined by evaluating the energy dissipated in the material during mechanical vibration. High damping materials, which have the ability to dissipate mechanical vibration energy, are valuable to be applied in the fields of noise control and in stabilizing structures in order to suppress mechanical vibrations and attenuate wave propagation [1-2]. Practical applications need low density materials that simultaneously exhibit high damping capacity and good ductility. However, in metals these properties are usually incompatible because of the dependence on microscopic mechanisms involved in strengthening and damping [3].

The compatibility of high damping capacity with high strength has been considered to be important for structural damping capacity of the aluminum severely deformed by materials subjected to resonance loading. However, increases in damping by various methods have been accompanied by decreases in strength [4]. It is established that severe plastic deformation is viable to produce high strength metals with ultrafine grained microstructure [5, 6].

The severe plastic deformation is reliable also as a process to produce high damping materials since the severely formed metals contain significant amount of lattice defects which give rise to damping capacity. Zheng Ming Yi et al. have reported the high damping capacity of Mg-Cu-Mn alloy severely deformed by equal channel angular press (ECAP) [7]. On the other hand, cold rolling is the one of the severe plastic deformation processes applicable to continuous production of large bulky materials [8, 9].

The increase of the damping capacity of the aluminium also can be achieved by the application of precipitation hardening treatment [10, 11] and superheating treatment [12]. Choong Do Lee reported that the precipitation hardening of coherent $Mg_2Si$ on T6 treatment in Al-7Si-0.3Mg alloy play a fundamental role in the simultaneous enhancement of mechanical property and damping capacity [10]. The purpose of this research is to study the effects of rolling deformation and annealing treatment on damping capacity of 1200 aluminium alloy.

**Experimental Procedure**

The raw material used was as-received 1200 aluminium alloys in sheet form with 1.3 mm thickness. The samples were cut into 70 mm length x 12 mm width dimensions for homogenization treatment at 560°C for 4 hours in a normal atmosphere and then cooled in the furnace to room temperature. Subsequently, the samples were undergone cold rolling process by using cold rolling machine to produce 20, 30 and 40 percent reductions respectively. Then, the samples were annealed at two different temperatures of 345°C and 400°C for 1 and 3 hours soaking times respectively.

A dynamic mechanical analyzer (Pyris Diamond DMA model, USA) was used to measure the damping capacity. Dynamic mechanical analysis was carried out in the three point bending mode using a dual cantilever system. The samples were prepared in the form of rectangular bars with dimensions of 50 x 10 x 1.0 mm. The tested specimens were run at 5°C/minute heating rate from 30 to 400ºC with 100 µm strain at 10 Hz vibration frequency in a flowing purified nitrogen gas.

The microstructures after homogenization process, rolling reduction and annealing process were analyzed by using optical microscope. Specimens were prepared by the standard metallography methods of cutting and mounting followed by wet grinding on a series of SiC papers. Finally, the specimens were polished with 6 µm, 3 µm and 1 µm diamond suspension using napless cloth. The etchant used was 1.0% HF in order to reveal the microstructures.

**Results and Discussion**

*i) Damping Capacity*

Figures 1, 2, 3 and 4 display the damping capacities as a function of temperature after homogenized (H) at 560°C for 4 hours, rolling at various percent of rolling reductions (RR) and followed by annealed (A) at 345°C (1 hour and 3 hours soaking times) and 400°C (1 hour and 3 hours soaking times) respectively and also homogenized sample.

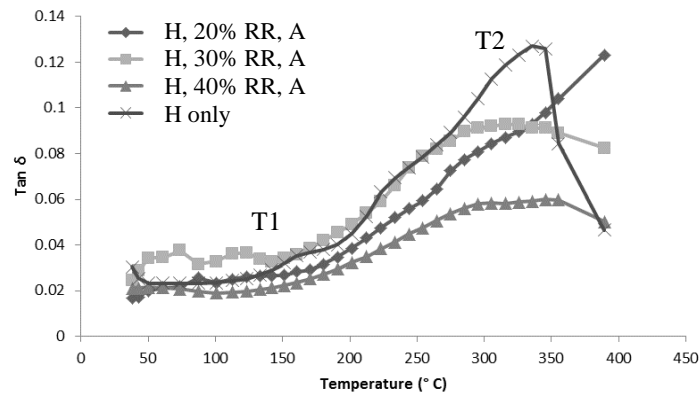

Figure 1 Damping capacities as a function of temperature for various percents of rolling reduction after annealed at 345°C for one hour soaking time and homogenized sample.

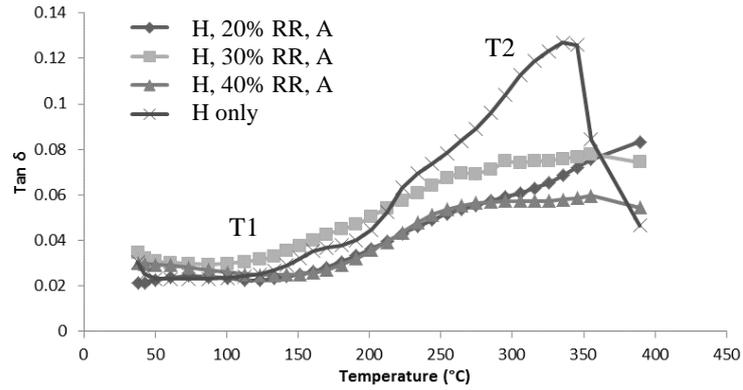

Figure 2  Damping capacities as a function of temperature for various percents of rolling reduction after annealed at 345°C for three hours soaking time and homogenized sample.

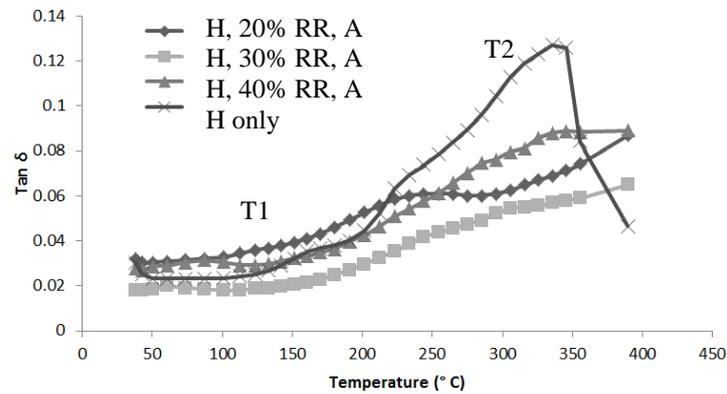

Figure 3  Damping capacities as a function of temperature for various percents of rolling reduction after annealed at 400°C for one hour soaking time and homogenized sample.

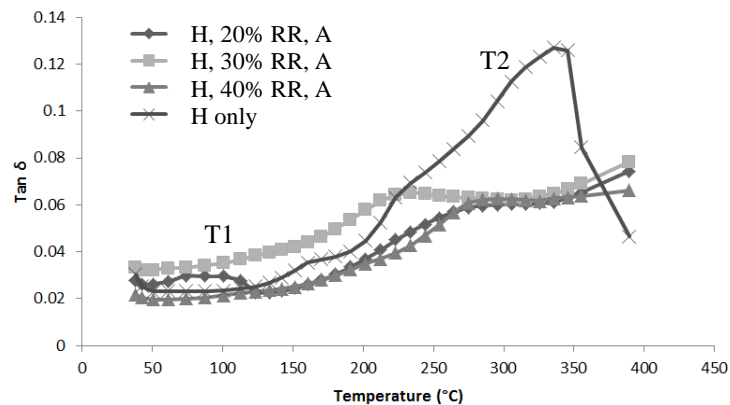

Figure 4  Damping capacities as a function of temperature for various percents of rolling reduction after annealed at 400°C for three hours soaking time and homogenized sample.

Basically, the damping capacity generally exhibited an increase at elevated temperatures for all samples. Two significant transition points were observed in Figure 1 and Figure 2 which termed as T1 and T2 respectively. T1 transition point took place at about 150°C in Figure 1 meanwhile in between 100 to 150°C in Figures 2, 3 and 4. In all Figures 1 to 4, T2 took place at about 340°C in homogenized samples meanwhile T2 took place at about 350°C in all homogenized, rolled reduction and annealed samples.

In general, 30 percent and 40 percent rolling reductions in Figures 1 and 2 started to decrease just after 350°C temperature. It have been reported that the grain boundary for high rolling reduction were difficult to slip at high temperature and cause the damping capacity decreases [13] and also believe to be depending also on the annealing parameters used. In this study, it also can be seen that homogenized sample cannot maintain the damping capacity and drop drastically after achieved the peak just after 350°C temperature. The same trend of decreasing of damping capacity in as-rolled magnesium sheets also have been observed after achieved the peak at around 225°C [14].

In this study, the promising results in the increasing of damping capacity values at higher temperatures (above 350°C) has been achieved by homogenization at 560°C for 4 hours followed by 20 percent rolling reduction annealing at 345°C for 1 hour relatively. Previous study by Ning Ma et al. have reported that the increase of damping capacity is attributed by the decreasing of annealing temperature [14]. However, in composite system combined with roll bonding, the results indicate that by increasing the percentage of reinforcing phase, the damping capacity increases. It is obvious that the damping capacity of the composite is higher than that of Al6061 alloy. Results show that the increase of damping by internal friction is due to the presence of SiC particles [3].

*ii) Microstructural Evolution*

The microstructural evolution of the surface layer along the rolling direction of 1200 aluminium alloy after homogenization, rolling reduction and annealing the processes are presented in Figure 5. From Figure 5, it could be said that the increase in percent reduction during cold rolling procedure has led to a decrease of the average grain size of surface layers of the alloy. It can be observed that when the rolling percent reduction was 40% (in Figure 5d), the equiaxed grains became smaller and denser.

Zhang et al. reported that high damping capacity has been achieved in ultrafine-grained pure aluminum L12 with a mean grain size of 1 μm was produced by equal channel angular pressing (ECAP) and annealing at 150°C for 2 hours [5]. Koizumi et al. also reported that pure aluminum that refined to ultrafine grains, smaller than 1 μm which possess a high damping capacity after accumulative roll bonding for five cycles [4].

**Conclusions**

The following conclusions can be drawn from this study:

i ) The decreasing of damping capacity in 30 percent and 40 percent reductions just after 350°C temperature was due to difficulty of grain boundary to slip at high temperature.
ii ) A lower annealing temperature at 375°C for 1 hour has produced increasing of damping capacity values after 350°C in 20 percent rolling reduction relatively.
iii) The combination of homogenization, rolling reduction and annealing processes is viable to be applied in order to sustain a stable damping capacity value at about 350°C in 1200 aluminium alloy. Homogenization at 560°C for 4 hours followed by 20 percent rolling reduction and annealing at 345°C for 1 hour is the optimal treatment for 1200 aluminium alloy.

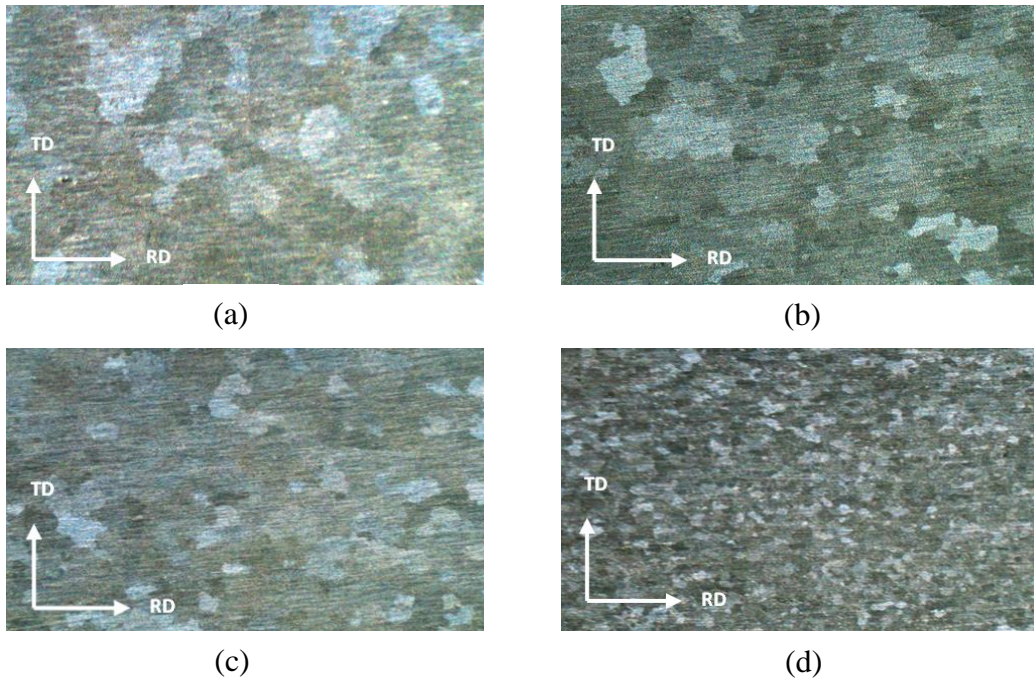

Figure 5 Optical micrographs show the microstructural evolution after homogenization, rolling reduction and annealing at 50x magnification of
(a) 20 percent reduction, 345°C for 1 hour.
(b) 40 percent reduction, 345°C for 1 hour.
(c) 20 percent reduction, 400°C for 3 hours.
(d) 40 percent reduction, 400°C for 3 hours.
*Note : TD is tranverse direction, RD is rolling direction.